\documentclass[nohyper,12pt,letterpaper]{JHEP3}
\usepackage{epsfig}

\author{Robert de Mello Koch$^{1,3}$, Jeff Murugan$^{2,4}$, Jelena Smolic$^{1}$ 
and Milena Smolic$^{1}$\\
${}^{1}$Department of Physics and Centre for Theoretical Physics,\\ 
University of the Witwatersrand, Wits, 2050, South Africa\\
\qquad\\
${}^{2}$Department of Physics, Brown University\\
Providence, RI02912 , USA \\
\qquad\\
${}^{3}$Stellenbosch Institute for Advanced Studies,\\
Stellenbosch, South Africa\\
\qquad\\
${}^4$Department of Mathematics and Applied Mathematics,\\
University of Cape Town, Private Bag, Rondebosch 7700, South Africa\\
\qquad\\
E-mail: \email{robert@neo.phys.wits.ac.za, jeff@now.het.brown.edu, smolicj@science.pg.wits.ac.za, msmolic@webmail.co.za}}

\abstract{
In this article we study a pp-wave limit of the Lunin-Maldacena background. We show that
the relevant string theory background is a homogeneous pp-wave. We obtain the string
spectrum. The dual field theory is a deformation of
${\cal N}=4$ super Yang-Mills theory. We have shown that, for a class of operators, at
$O(g_{YM}^2)$ and at leading order in $N$, all contributions to the anomalous dimension
come from $F$-terms. We are able to identify the operator in the
deformed super Yang-Mills which is dual to the lowest string mode. By studying the
undeformed theory we are able to provide some evidence, directly in the field theory,
that a small set of nearly protected operators decouple.
We make some comments on operators in the Yang-Mills theory that are dual to excited 
string modes.}

\preprint{Brown-Het-1446\\ Wits-CTP-022}

\title{Deformed PP-waves from the Lunin-Maldacena Background}


\def \Tr{\mbox{Tr\,}}

\begin{document}

\section{Introduction}
The AdS/CFT correspondence\cite{AdSCFT} relates string theory on 
negatively curved spacetime and large $N$ quantum field theories. 
The correspondence is a strong/weak coupling duality in the 't Hooft 
coupling of the field theory. At large $N$ and large 't Hooft coupling, 
both quantum gravity and curvature corrections in the string theory
are supressed. The dual gauge theory however is strongly coupled.
For small 't Hooft coupling and large $N$, the gauge theory coupling
is small, but curvature corrections in the string 
theory are not. Computations that can be carried out on 
both sides of the correspondence necessarily involve quantities that 
are not corrected or receive small corrections, allowing weak coupling 
results to be extrapolated to strong coupling. A very interesting class
of observables of this type are the near BPS operators discovered by
Berenstein, Maldacena and Nastase\cite{BMN}, which are dual to excited string modes.
Indeed, since these operators are not BPS, the BMN limit of the gauge 
theory reproduces genuinely stringy physics, via the AdS/CFT correspondence.

In a recent article\cite{LM}, Lunin and Maldacena (LM) have studied $\beta$-deformations
of the ${\cal N}=4$ super Yang-Mills theory, and have identified the
corresponding gravitational deformation of the AdS$_5\times$S$^5$ background. 
The field theory deformation is obtained by
making the following replacement in the superpotential

\begin{equation}
\Tr \left(\hat{\Phi}^1\hat{\Phi}^2\hat{\Phi}^3 -
\hat{\Phi}^1\hat{\Phi}^3\hat{\Phi}^2\right)\to
\Tr \left(e^{i\pi\gamma}\hat{\Phi}^1\hat{\Phi}^2\hat{\Phi}^3 -
e^{-i\pi\gamma}\hat{\Phi}^1\hat{\Phi}^3\hat{\Phi}^2\right).
\label{fdef}
\end{equation}

\noindent
The deformed field theory has ${\cal N}=1$ supersymmetry and is invariant
under a $U(1)\times U(1)$ non-${\cal R}$ symmetry. The ${\cal N}=4$ theory is dual 
to string theory in the AdS$_5\times$S$^5$ geometry, which contains a two torus. 
The isometries of the two torus match with the $U(1)\times U(1)$ field theory symmetry.
Denote the metric of this two torus by $g$ and the NS-NS two form (which is of 
course zero in the undeformed theory) by $B$. The deformation of the dual gravitational 
theory is obtained by replacing

$$\tau =B+i\sqrt{g}\to \tau_\gamma ={\tau\over 1+\gamma\tau}.$$

\noindent
The AdS$_5$ factor is unchanged which is expected because (\ref{fdef}) is a marginal
deformation. Studies of the AdS/CFT correspondence for this deformation are likely
to produce interesting results for at least two reasons. Firstly, it is important to
generalize the AdS/CFT correspondence to less supersymmetric examples. Secondly, since
this background has a continuous adjustable deformation parameter, it may be possible to
define new scaling limits.

A study of semiclassical string states provided important insights into the BMN 
limit\cite{GKP}. Motivated by this, semiclassical string states in the LM background 
were recently compared to a class of gauge theory scalar operators\cite{FRT}. 
The 1-loop anomalous dimensions of these operators are described by an integrable 
spin chain and match beautifully with the energies of the semiclassical string states. 
Further, by employing the Lax pair for strings
in the LM background\cite{F}, the Landau-Lifschitz action associated to the one-loop spin
chain was recovered. This indicates that the integrable structures in the gauge 
theory and the string theory match. Further analysis of the relevant spin chain is
given in \cite{Beisert}. For further recent insights into the gauge/string
correspondence for these (and other) new examples see\cite{BK}.

The logic employed by Lunin and Maldacena to obtain the gravitational theory dual to the
deformed field theory can be extended in a number of ways. Recently, instead of deforming
the ${\cal N}=4$ super Yang-Mills theory, deformations of ${\cal N}=1$ and ${\cal N}=2$
theories have been considered\cite{GN}. Further, deformations of eleven
dimensional geometries of the form AdS$_4\times$Y$_7$ with $Y_7$ a seven dimensional
Sasaki-Einstein\cite{CA},\cite{Lee} or weak $G_2$ or tri-Sasakian\cite{Lee} 
space have been considered. 

In this article we are interested in studying a pp-wave limit of the LM background.
There are a number of interesting pp-wave limits that can be taken. Each of these limits
allows us to probe different {\it stringy} aspects of the correspondence, and are thus
worthy of study. One such limit was in fact already considered in \cite{NP}. 
We will be considering a different pp-wave 
limit, to provide further independent support for and insight into the correspondence
of \cite{LM}.

The paper is organized as follows: in section 2 we describe the pp-wave limit that we
consider here. We obtain the metric and B field by taking an appropriate limit of the
results in \cite{LM}. The resulting background is a homogeneous 
pp-wave\cite{HPP}. It is known that the 
string sigma model in this background can be solved exactly\cite{SHPP}. We 
provide this analysis in
section 3. In section 4 we study the dual gauge theory and consider the question of 
how to define the
near-BPS operators with anomalous dimensions which reproduce the spectrum of the string
sigma model. We are able to argue that, in the large $N$ limit and at one loop, one can 
ignore gluon exchange, self energy insertions and $D$-term contributions. This allows
a significant simplification of the analysis. We are able to identify the operator dual
to the lowest string mode. For $\gamma=0$ we are able to find near-BPS operators which 
reproduce the spectrum of the string sigma model. Our results are consistent with the
expected decoupling of a small set of nearly protected operators.
In the $\gamma\ne 0$ case, we are able argue that there is a set of nearly protected
operators whose spectrum of anomalous dimensions is independent of $\gamma$ in agreement
with the string theory result. We also find near BPS operators for small values of the 
${\cal R}$ charge $J$ and find that their anomalous dimension does depend on $\gamma$.
Section 5 is reserved for a discussion of our results.

\section{PP-wave Limit of the Lunin-Maldacena Geometry}

In this section we will take the pp-wave limit of the LM background. Our goal is to
obtain the spectrum of free strings in this background. To write down the relevant
string sigma model, we need only the metric and the B field. Thus, we do not consider
the RR-fluxes $C_2$ and $C_4$ which are also non-zero in the LM background.

The metric is\cite{LM}

\begin{eqnarray}
\nonumber
ds^2 &=&R^2\left(
-dt^2\cosh^2\rho +d\rho^2 +\sinh^2\rho d\Omega_3^2+
\sum_i d\mu_i^2\right.\\
&+& \left. G\sum_{i=1}^3\mu_i^2 d\phi_i^2
+\gamma^2 \mu_1^2\mu_2^2\mu_3^2 G\left(\sum_i d\phi_i\right)^2
\right).
\end{eqnarray}

\noindent
where

$$\mu_1=\cos\alpha,\qquad\mu_2 =\sin\alpha\cos\theta,\qquad
\mu_3 =\sin\alpha\sin\theta,$$
$$G={1\over 1+\gamma^2 
(\mu_1^2\mu_2^2 +\mu_1^2\mu_3^2 +\mu_3^2\mu_2^2)}.$$

\noindent
It is useful to use the angles $\psi$, $\varphi_1$ and $\varphi_2$, defined by

$$\phi_1 =\psi-\varphi_2,\qquad\phi_2=\psi+\varphi_1+\varphi_2,\qquad
\phi_3=\psi-\varphi_1 .$$

\noindent
The parameter $\gamma$ is the deformation parameter. We will perform the 
Penrose limit using the null geodesic $\tau =\psi$, with 
$\alpha_0=\cos^{-1}{1\over\sqrt{3}}$ and $\theta_0={\pi\over 4}.$ We set

$$\theta ={\pi\over 4}+\sqrt{2\over 3}{x^1\over R} ,\qquad 
\alpha =\alpha_0 -{x^2\over R},\qquad 
\rho={r\over R}$$
$$\varphi_1={x^{3\prime}\over R},\qquad 
\varphi_2={x^{4\prime}\over R},\qquad
t=x^+ +{x^- \over R^2},\qquad \psi ={x^- \over R^2}-x^+ , $$

$$ x^3=\sqrt{2\over 3+\gamma^2}\left(x^{3\prime}+{1\over 2}x^{4\prime}\right),\qquad
x^4=\sqrt{3\over 2(3+\gamma^2)}x^{4\prime},$$

\noindent
and take the limit $R\to\infty$. The pp-wave metric we obtain is

\begin{eqnarray}
\nonumber
ds^2 &=& -4dx^+ dx^- -\left[r^2+{4\gamma^2\over 3+\gamma^2}
\left((x^1)^2 + (x^2)^2\right)\right](dx^+)^2 +dr^2 +r^2 d\Omega_3^2\\
&+&(dx^1)^2+(dx^2)^2+ (dx^3)^2 +(dx^4)^2 
+{4\sqrt{3}\over\sqrt{3+\gamma^2}}\left(x^1dx^3+x^2 dx^4
\right) dx^+.
\end{eqnarray}

To obtain the string sigma model, we will also need the $B$ field in the pp-wave limit.
We find

$$B_{\varphi_1\varphi_2}{\cal D}\varphi_1\wedge{\cal D}\varphi_2=
G\gamma gR^2{\cal D}\varphi_1\wedge{\cal D}\varphi_2,$$

\noindent
where

$${\cal D}\varphi_1=d\varphi_1-d\psi +{3\mu_1^2\mu_2^2\over g}d\psi, \qquad
{\cal D}\varphi_2=d\varphi_2-d\psi +{3\mu_3^2\mu_2^2\over g}d\psi.$$

\noindent
Taking the pp-wave limit as above, we find the following $B$ field

$$ B={\gamma\over\sqrt{3}}d x^3\wedge d x^4 
+{2\gamma\over\sqrt{3+\gamma^2}}
(x^2 d x^3\wedge dx^+ + x^1 dx^+\wedge dx^4 ),$$

\noindent
and the following field strengths

$$ H_{23+}={2\gamma\over\sqrt{3+\gamma^2}},
\qquad H_{14+}=-{2\gamma\over\sqrt{3+\gamma^2}}.$$

\noindent
Thus, the field strength is {\it null} as it should be in the pp-wave limit.

\section{Strings in the PP-wave Limit of the Lunin-Maldacena Geometry}

Given the metric and $B$ fields written down in the previous section, in this
section we consider the resulting string sigma model. We show that this background
corresponds to a homogeneous pp-wave\cite{HPP} and are thus able to use existing
results\cite{SHPP} to obtain the spectrum.

We will be working in lightcone gauge. The string worldsheet action is
(we are dropping the fermions from our analysis)
$$S=-{1\over 4\pi\alpha'}\int d^2\sigma \left[\sqrt{\eta}\eta^{ab}g_{\mu\nu}
\partial_a X^\mu\partial_b X^\nu +\epsilon^{ab}B^{NS}_{\mu\nu}\partial_a X^\mu
\partial_b X^\nu +\alpha'\sqrt{\eta}\phi(x)R\right],$$

\noindent
with $R$ the scalar curvature on the worldsheet, $\eta^{ab}$ is the worldsheet
metric and $\eta=|\det {\eta_{ab}}|$.
We will choose $\sqrt{\eta}\eta^{ab}$ diagonal with $\sqrt{\eta}\eta^{00}=-1$ and 
$\sqrt{\eta}\eta^{11}=1$. After shifting

$$ x^- \to x^-+{\sqrt{3}\over 2\sqrt{3+\gamma^2}}
(x^1x^3+x^2x^4),$$

\noindent
the metric becomes

\begin{eqnarray}
ds^2 &=&
-4dx^+ dx^- -\left[\sum_{i=5}^8(x^i)^2+{4\gamma^2\over 3+\gamma^2}
\left((x^1)^2 + (x^2)^2\right)
\right](dx^+)^2 \nonumber\\
&+&\sum_{i=1}^8 (dx^i)^2
+{2\sqrt{3}\over\sqrt{3+\gamma^2}}\left(x^1dx^3 -x^3dx^1 +x^2 dx^4-x^4dx^2
\right) dx^+ .\nonumber
\end{eqnarray}

\noindent
This metric corresponds to a homogeneous pp-wave\cite{HPP}. The sigma model for this
background has been considered in \cite{SHPP}; we will review the relevant results here.
In the gauge $x^+ =\tau$, we obtain the following Lagrangian density (we take $\sigma$ to
run from $0$ to $\pi$ and set $\alpha'={1\over 2\pi}$)\footnote{We thank
T. Mateos for pointing out an error in the next formula, which appeared in an 
earlier version of this draft.}

\begin{eqnarray}
{\cal L}
&=&-2{\partial x^-\over\partial \tau}
-{1\over 2}\left[\sum_{i=5}^8(x^i)^2+{4\gamma^2\over 3+\gamma^2}
((x^1)^2+(x^2)^2)\right]+{2\gamma\over\sqrt{3+\gamma^2}}\left(
x^2{\partial x^3\over\partial\sigma}-x^1{\partial x^4\over\partial\sigma}\right)
\nonumber \\
\nonumber
&+&{\sqrt{3}\over \sqrt{3+\gamma^2}}\left(
x^1{\partial x^3\over\partial\tau}-x^3{\partial x^1\over\partial\tau}+
x^2{\partial x^4\over\partial\tau}-x^4{\partial x^2\over\partial\tau}
\right)-{1\over 2}\sum_{i =1}^8\partial_a x^i\partial^a x^i .
\end{eqnarray}

\noindent
To quantize the theory, compute the canonical momenta

$$p^1(\tau,\sigma)=\dot{x}^1 (\tau,\sigma)
-\sqrt{3 \over 3+\gamma^2}x^3,\qquad
p^2(\tau,\sigma)=\dot{x}^2 (\tau,\sigma)
-\sqrt{3 \over 3+\gamma^2}x^4,$$
$$p^3(\tau,\sigma)=\dot{x}^3 (\tau,\sigma)
+\sqrt{3 \over 3+\gamma^2}x^1,\qquad
p^4(\tau,\sigma)=\dot{x}^4 (\tau,\sigma)
+\sqrt{3 \over 3+\gamma^2}x^2,$$
$$p^k(\tau,\sigma)={\partial {\cal L}\over\partial \dot{x}^k}=\dot{x}^k (\tau,\sigma)
\qquad k=5,6,7,8,$$

\noindent
and impose the equal time commutation relations

$$\big[p^k(\tau,\sigma),x^j(\tau,\sigma')\big]=-i\delta^{jk}\delta (\sigma-\sigma').$$

\noindent
The Hamiltonian is 

\begin{eqnarray}
H={1\over 2}\int_0^\pi d\sigma \left[\sum_{k=1}^8
\left(p^k p^k+{\partial x^k\over\partial\sigma}{\partial x^k\over\partial\sigma}\right)
+{3\over 3+\gamma^2}\sum_{k=1}^4 (x^k)^2+\sum_{k=5}^8(x^k)^2 +
{4\gamma^2\over 3+\gamma^2}\sum_{k=1}^2 (x^k)^2\right.\nonumber\\
\left. -{4\gamma\over\sqrt{3+\gamma^2}}\left(x^2{\partial x^3\over\partial\sigma}
-x^1{\partial x^4\over\partial\sigma}\right)+{2\sqrt{3}\over\sqrt{3+\gamma^2}}
\left(p^1x^3+p^2x^4-p^3x^1-p^4x^2\right)\right] .\nonumber
\end{eqnarray}

\noindent
Notice that the modes corresponding to $x^5,x^6,x^7,x^8$ have masses that do not depend
on $\gamma$, i.e. they are unaffected by the deformation. This is not unexpected, since
these coordinates come from the AdS$_5$ part of the space which does not participate in
the deformation. We will, from this point on, consider only $x^1,x^2,x^3,x^4$.

The Heisenberg equations of motion are

$${\partial^2 x^i\over\partial t^2}-{\partial^2 x^i\over\partial \sigma^2}
+f^{ij}{\partial x^j\over\partial t}+h^{ij}{\partial x^j\over\partial \sigma}
+k_i x^i=0,$$

\noindent
where

$$f^{ij}=\left[\matrix{
0 &0 &-2\sqrt{3\over 3+\gamma^2} &0\cr 
0 &0 &0 &-2\sqrt{3\over 3+\gamma^2}\cr 
2\sqrt{3\over 3+\gamma^2} &0 &0 &0\cr 
0 &2\sqrt{3\over 3+\gamma^2} &0 &0}\right],$$

\noindent
and

$$h^{ij}=2\left[\matrix{
0 &0 &0 &{\gamma\over \sqrt{3+\gamma^2}}\cr 
0 &0 &-{\gamma\over \sqrt{3+\gamma^2}} &0\cr 
0 &{\gamma\over \sqrt{3+\gamma^2}} &0 &0\cr 
-{\gamma\over \sqrt{3+\gamma^2}} &0 &0 &0}\right],$$
$$k_1=k_2={4\gamma^2\over 3+\gamma^2},\qquad k_3=k_4=0.$$

\noindent
To solve these equations introduce the mode expansions

$$ x^i(t,\sigma)=\sum_{n=-\infty}^\infty x_n^i (t)e^{2in\sigma}.$$

\noindent
Reality of $x^i(t,\sigma)$ is encoded (as usual) in

$$ x_n^i=(x_{-n}^i)^*.$$

\noindent
The equations of motion now become the following equation for the modes

$${\partial^2 x^i_n\over\partial t^2}+4n^2 x^i_n 
+f^{ij}{\partial x^j_n\over\partial t}+2in h^{ij} x^j_n 
+k_i x^i_n=0.$$

\noindent
Following \cite{SHPP} we now make the following ansatz

$$x_n^i(t)=a^{(n)}_j A^{(n)}_{ij}e^{i\omega^{(n)}_j t}.$$

\noindent
$a^{(n)}_j$ will be a destruction/annihilation operator; $A^{(n)}_{ij}$ is a unitary transformation
diagonalizing the equation of motion; $\omega^{(n)}_j$ is our spectrum. Plugging this into
the equations of motion we find

$$\left(-(\omega_k^{(n)})^2\delta^{ij}+4n^2\delta^{ij}+if^{ij}\omega_k^{(n)}+2inh^{ij}+k_i\delta^{ij}
\right)a_k^{(n)}A_{jk}^{(n)}e^{i\omega_k^{(n)}t}=0.$$

\noindent
The condition for a nontrivial solution is

$$\det \left(-(\omega_k^{(n)})^2\delta^{ij}+4n^2\delta^{ij}+if^{ij}\omega_k^{(n)}+2inh^{ij}+k_i\delta^{ij}
\right)=0,$$

\noindent
which leads to the following quartic equation

$$\omega^4 -(4+8n^2)\omega^2 +16n^4 =0.$$

\noindent
It is solved by

$$\omega
=1\pm\sqrt{1+4n^2}.$$

\noindent
This is in perfect agreement with \cite{Toni}. Notice that the spectrum is independent
of the deformation parameter $\gamma$. The fact that the spectrum is independent
of $\gamma$ is unexpected. Evidently, the $\gamma$ dependence in the $B$ field exactly
compensates for the $\gamma$ dependence of the geometry. 

\section{Dual Field Theory Analysis}

In this section, we will study the field theory obtained after deforming 
the superpotential (fields with a hat $\hat\Phi$ denote superfields; fields without a
hat $\Phi$ denote the Higgs fields - the bosonic bottom component of $\hat\Phi$)

$$ \Tr \left(\hat{\Phi}^1\hat{\Phi}^2\hat{\Phi}^3 -
\hat{\Phi}^1\hat{\Phi}^3\hat{\Phi}^2\right)\to
\Tr \left(e^{i\pi\gamma}\hat{\Phi}^1\hat{\Phi}^2\hat{\Phi}^3 -
e^{-i\pi\gamma}\hat{\Phi}^1\hat{\Phi}^3\hat{\Phi}^2\right).$$

\noindent
We consider only the Higgs fields. Our goal is to construct
operators dual to the string modes 
discussed in section 3.1; these will be built from the Higgs fields. The kinetic
terms and $D$ terms for the Higgs fields are invariant under the deformation.
The usual $F$ terms are however now replaced by

$$V=\Tr( \Big|\big[\Phi^2 ,\Phi^3\big]_\gamma\Big|^2
+ \Big|\big[ \Phi^3,\Phi^1\big]_\gamma\Big|^2 
+ \Big|\big[ \Phi^1,\Phi^2\big]_\gamma\Big|^2 ),$$

\noindent
where

$$\big[A ,B\big]_\gamma\equiv e^{i\pi\gamma}AB -e^{-i\pi\gamma}BA .$$

\noindent
In the undeformed theory\cite{Motl}, when computing correlators of traces
in the case that each trace 
involves only $\Phi^1$, $\Phi^2$ and $\Phi^3$ or 
$\bar\Phi^1$, $\bar\Phi^2$ and $\bar\Phi^3$, 
one does not need to consider $D$-term contributions, self energy corrections
or gluon exchange at order $g_{YM}^2$ in Yang-Mills perturbation theory. (See
\cite{Penati} for useful superspace techniques.) We
will argue that this is also true in the deformed theory, at leading order in $N$. 
Using this insight,
we construct operators in the Yang-Mills theory that are dual to the vacuum of
the sigma model. Next we study operators dual to excited string 
states in the undeformed ($\gamma=0$) theory. Finally, we reconsider 
this question in the deformed theory.

\subsection{Only $F$-terms contribute}

In this section we will consider correlators of the form

$$\langle {\cal O}\bar{\cal O}\rangle,$$

\noindent
where ${\cal O}$ is a trace of $k$ Higgs fields

$${\cal O}=f_{i_1 i_2\cdots i_k}\Tr \left( \Phi^{i_1}\Phi^{i_2}\cdots
\Phi^{i_k}\right),$$

\noindent
and the indices $i_1,...,i_k\in \{1,2,3\}$. We do not assume anything about 
the coefficient $f_{i_1 i_2\cdots i_k}$.
In the undeformed theory\cite{Motl}, one argues that the $D$-terms, gluon exchange 
and self energy corrections are all flavor blind at one-loop. Consequently, if we
are working to one loop order, we could replace

$${\cal O}\to\Tr \left((\Phi^1)^k\right).$$

\noindent
The result of \cite{HFS} tells us that the correlator 
$\langle\Tr \left((\Phi^1)^k\right)\Tr \left((\bar\Phi^1)^k\right)\rangle$ receives no
radiative corrections at $O(g_{YM}^2)$, so the result follows.

When we deform the theory, the $D$-term contributions and gluon exchange contributions
are unchanged. $F$-term contributions to the self energy need to be considered
carefully because the $F$-terms are affected by the deformation. The $F$-terms can be 
split into two pieces

$$ V_F = V_{inv}+V_{def},$$

\noindent
where

$$V_{inv}= 2\Tr (\Phi^1\Phi^2\bar\Phi^2\bar\Phi^1 + \Phi^2\Phi^1\bar\Phi^1\bar\Phi^2),$$
$$V_{def}=-2\Tr ( e^{-2\pi i\gamma}\Phi^2\Phi^1\bar\Phi^2\bar\Phi^1 + 
e^{2\pi i\gamma}\Phi^1\Phi^2\bar\Phi^1\bar\Phi^2).$$

\noindent
The self energy contribution coming from $V_{inv}$ will be the same as in the 
undeformed theory; self energy contribution coming from $V_{def}$ will not. The
Feynman diagrams corresponding to self energy contributions, coming from these
two vertices, are shown below. (A) shows the contribution from $V_{inv}$ and
(B) the contribution from $V_{def}$. Since (B) is a non-planar diagram, it can 
be dropped at large $N$ and consequently, the only contribution to the self energy
coming from the $F$-terms is invariant under the deformation to leading order in $N$.
Thus, for the correlators that we are considering, one does not need to consider $D$-term
contributions, 
self energy corrections or gluon exchange, at order $g_{YM}^2$ in Yang-Mills perturbation 
theory and at leading order in $N$. 

\begin{center}
\begin{figure}[h]{\psfig{file=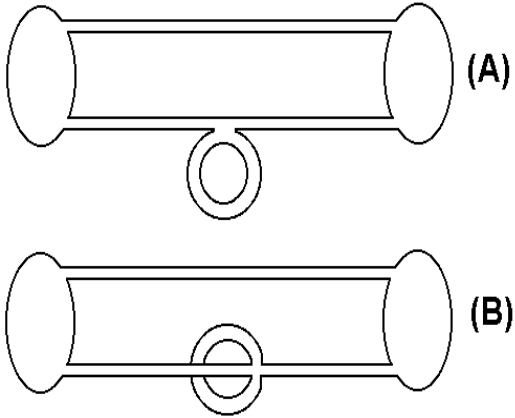,width=7cm,height=6cm}
 \caption{This plot shows the Feynman diagrams corresponding to self energy 
contributions coming from the $F$-terms. (A) shows the contribution from $V_{inv}$ and
(B) the contribution from $V_{def}$. (A) is $O(g_{YM}^2 N^3)$; (B) is $O(g_{YM}^2 N)$.}}
\end{figure}
\end{center}

\subsection{Operators dual to the Vacuum}

The operator dual to the vacuum of the string sigma model is a BPS operator.
Thus, we expect that the $U(1)_{\cal R}$ charge of this operator is equal to
its conformal dimension, and further that it is not charged under the $U(1)\times U(1)$
symmetry of the field theory. This follows because our pp-wave limit is taken by boosting
along $\psi$; there is no momentum in the $\varphi_1,\varphi_2$ directions.
The charges and dimension of the three Higgs fields are

$$\matrix{ &U(1) &U(1) &U(1)_{\cal R}\equiv J &\Delta\cr
\Phi^1 &0 &-1 &1 &1\cr
\Phi^2 &1 &1 &1 &1\cr
\Phi^3 &-1 &0 &1 &1}$$

\noindent
We will explicitly construct the operator dual to the vacuum for small values of
$J$. This will allow us to extract a rule that gives the correct operator for all $J$.

For $J=3$, there are two independent loops out of which the operator dual to the 
vacuum could be constructed

$${\cal O}_1=\Tr (\Phi^1\Phi^2\Phi^3),\qquad {\cal O}_2 =\Tr (\Phi^1\Phi^3\Phi^2).$$

\noindent
Using the two point function of the Higgs fields 
(indices $a,b,c,d=1,...,N$ are color labels;
indices $j,k=1,2,3$)

$$\langle \bar\Phi^j_{ab}(x)\Phi^k_{cd}(0)\rangle =\delta^{jk}\delta_{ad}\delta_{bc}
{1\over 4\pi^2 |x|^2}\equiv \delta^{jk}\delta_{ad}\delta_{bc}{a\over |x|^2},$$

\noindent
we compute the planar contribution to

\begin{equation}
\langle\bar{\cal O}_i(x_1) V_F(y){\cal O}_j(x_2)\rangle =
{\cal M}_{ij} {a^5\over |x_1-y|^4 |x_2-y|^4 |x_1-x_2|^2}N^4.
\label{keycor}
\end{equation}

\noindent
We use the above correlator to define the matrix ${\cal M}$.
The result is

$${\cal M}^T =
\left[\matrix{6 &-6e^{-2\pi i\gamma}\cr -6e^{2\pi i\gamma} &6}\right].$$

\noindent
The matrix ${\cal M}$ has a single zero eigenvalue. The operator dual to
the vacuum is given by that linear combination which corresponds to the
zero eigenvalue - it is the two point function of this linear combination 
that is not corrected, as expected for a BPS operator. There is a single zero 
eigenvalue which implies that this state is unique. Our operator is

$$ {\cal O}_\gamma =\Tr\left( \Phi^1\Phi^2\Phi^3\right)+e^{-2\pi i\gamma}\Tr \left(
\Phi^1\Phi^3\Phi^2\right).$$

\noindent
It has dimension $\Delta =3$, $U(1)_{\cal R}$ charge $J=3$ and is neutral under
$U(1)\times U(1)$. Notice that when $\gamma=0$ our deformed operator recovers the
expected BPS operator of the undeformed case and further that ${\cal O}_\gamma $ respects 
the $Z_3$ symmetry (which acts as a cyclic permutation of the three Higgs fields) of the 
deformed theory.

Next, consider $J=6$. In this case, our operator is a linear combination of 16 loops

$${\cal O}_1=\Tr\left(\Phi^1 \Phi^1 \Phi^2 \Phi^2 \Phi^3 \Phi^3\right)\,\,
{\cal O}_2=\Tr\left(\Phi^1 \Phi^1 \Phi^2 \Phi^3 \Phi^2 \Phi^3\right)\,\,
{\cal O}_3=\Tr\left(\Phi^1 \Phi^1 \Phi^2 \Phi^3 \Phi^3 \Phi^2\right)$$
$${\cal O}_4=\Tr\left(\Phi^1 \Phi^1 \Phi^3 \Phi^2 \Phi^2 \Phi^3\right)\,\,
{\cal O}_5=\Tr\left(\Phi^1 \Phi^1 \Phi^3 \Phi^2 \Phi^3 \Phi^2\right)\,\,
{\cal O}_6=\Tr\left(\Phi^1 \Phi^1 \Phi^3 \Phi^3 \Phi^2 \Phi^2\right)$$
$${\cal O}_7=\Tr\left(\Phi^1 \Phi^2 \Phi^1 \Phi^2 \Phi^3 \Phi^3\right)\,\,
{\cal O}_8=\Tr\left(\Phi^1 \Phi^2 \Phi^1 \Phi^3 \Phi^2 \Phi^3\right)\,\,
{\cal O}_9=\Tr\left(\Phi^1 \Phi^2 \Phi^1 \Phi^3 \Phi^3 \Phi^2\right)$$
$${\cal O}_{10}=\Tr\left(\Phi^1 \Phi^2 \Phi^2 \Phi^1 \Phi^3 \Phi^3\right)\,\,
{\cal O}_{11}=\Tr\left(\Phi^1 \Phi^2 \Phi^2 \Phi^3 \Phi^1 \Phi^3\right)\,\,
{\cal O}_{12}=\Tr\left(\Phi^1 \Phi^2 \Phi^3 \Phi^1 \Phi^2 \Phi^3\right)$$
$${\cal O}_{13}=\Tr\left(\Phi^1 \Phi^2 \Phi^3 \Phi^1 \Phi^3 \Phi^2\right)\,\,
{\cal O}_{14}=\Tr\left(\Phi^1 \Phi^2 \Phi^3 \Phi^2 \Phi^1 \Phi^3\right)\,\,
{\cal O}_{15}=\Tr\left(\Phi^1 \Phi^3 \Phi^1 \Phi^3 \Phi^2 \Phi^2\right)$$
$${\cal O}_{16}=\Tr\left(\Phi^1 \Phi^3 \Phi^2 \Phi^1 \Phi^3 \Phi^2\right).$$

\noindent
These operators were selected by requiring that they have $\Delta =J=6$,
and zero $U(1)\times U(1)$ charge. We again want to identify the linear combination 
of these operators that is BPS. As for the case $J=3$, we do this by looking for the 
linear combination whose two point function does not receive corrections 
at $O(g_{YM}^2)$ and at leading order in $N$. By studying the correlator (\ref{keycor})
we can read off ${\cal M}$; null vectors of ${\cal M}$ are then natural candidate
BPS operators. In this case, again at leading order in $N$ we find

$${\cal M}^T=2\left[
\matrix{
3   &b   &0   &0   &0   &0 &b   &0   &0   &0   &b   &0   &0   &0   &0   &0\cr
b^* &5   &b   &b   &0   &0 &0   &b   &0   &0   &0   &0   &0   &b   &0   &0\cr
0   &b^* &4   &0   &b   &0 &b^* &0   &b   &0   &0   &0   &0   &0   &0   &0\cr
0   &b^* &0   &4   &b   &0 &0   &0   &0   &0   &b^* &0   &0   &0   &b   &0\cr
0   &0   &b^* &b^* &5   &b &0   &b^* &0   &0   &0   &0   &0   &b^* &0   &0\cr
0   &0   &0   &0   &b^* &3 &0   &0   &b^* &0   &0   &0   &0   &0   &b^* &0\cr
b^* &0   &b   &0   &0   &0 &5   &b   &0   &b   &0   &0   &b   &0   &0   &0\cr
0   &b^* &0   &0   &b   &0 &b^* &6   &b   &0   &0   &b^* &0   &0   &0   &b\cr
0   &0   &b^* &0   &0   &b &0   &b^* &5   &b^* &0   &0   &b^* &0   &0   &0\cr
0   &0   &0   &0   &0   &0 &b^* &0   &b   &4   &b^* &0   &0   &0   &b   &0\cr
b^* &0   &0   &b   &0   &0 &0   &0   &0   &b   &5   &0   &b   &b   &0   &0\cr
0   &0   &0   &0   &0   &0 &0   &2b  &0   &0   &0   &6   &2b  &2b   &0   &0\cr
0   &0   &0   &0   &0   &0 &b^* &0   &b   &0   &b^* &b^* &6   &0   &b   &b\cr
0   &b^* &0   &0   &b   &0 &0   &0   &0   &0   &b^* &b^* &0   &6   &b   &b\cr
0   &0   &0   &b^* &0   &b &0   &0   &0   &b^* &0   &0   &b^* &b^* &5   &0\cr
0   &0   &0   &0   &0   &0 &0   &2b^*   &0   &0   &0   &0   &2b^* &2b^* &0   &6}
\right]$$

\noindent
where $b=-e^{-2\pi i\gamma}$ and $b^*=-e^{2\pi i\gamma}$.
Again, ${\cal M}$ has a single zero eigenvalue, so that there is an
operator whose two point function does not get corrected at $O(g_{YM}^2)$ 
and it is again unique. For $J=9$ there are 188 basis loops. In this case 
${\cal M}$ again has a single null vector, so that we again have a unique candidate
BPS operator.

By studying the candidate BPS operators for $J=3,6,9$ we have been able to
identify a rule which allows us to write down a candidate BPS operator for
{\it any} $J$. To write down our rule, we call the following exchanges

$$\Phi^1\Phi^2\to\Phi^2\Phi^1,\quad {\rm or}\quad 
  \Phi^2\Phi^3\to\Phi^3\Phi^2,\quad {\rm or}\quad
  \Phi^3\Phi^1\to\Phi^1\Phi^3 ,$$

\noindent
{\it even exchanges} and the exchanges

$$\Phi^2\Phi^1\to\Phi^1\Phi^2,\quad {\rm or}\quad 
  \Phi^3\Phi^2\to\Phi^2\Phi^3,\quad {\rm or}\quad
  \Phi^1\Phi^3\to\Phi^3\Phi^1 ,$$

\noindent
{\it odd exchanges}. Consider for illustration the case with $J=6$.
To construct the vacuum state, we start from the
loop $\Tr\left(\Phi^1 \Phi^1 \Phi^2 \Phi^2 \Phi^3 \Phi^3\right)$ and perform a sequence
of even and odd exchanges until we generate the full 16 operators generated above.
For each odd exchange we append the factor $\alpha^*=e^{2\pi\gamma i}$, and for each even
exchange we append the factor $\alpha =e^{-2\pi\gamma i}$. Thus, for example, since

$$\Tr\left(\Phi^1 \Phi^1 \Phi^2  \Phi^3 \Phi^3\Phi^2\right)$$

\noindent
is obtained from

$$\Tr\left(\Phi^1 \Phi^1 \Phi^2 \Phi^2 \Phi^3 \Phi^3\right)$$

\noindent
by performing two even exchanges

$$\Tr\left(\Phi^1 \Phi^1 \Phi^2 \Phi^2 \Phi^3 \Phi^3\right)\to
\Tr\left(\Phi^1 \Phi^1 \Phi^2  \Phi^3 \Phi^2\Phi^3\right)\to
\Tr\left(\Phi^1 \Phi^1 \Phi^2\Phi^3 \Phi^3\Phi^2\right),$$

\noindent
we know that it will have a phase of $\alpha^2 =(e^{-2\pi i\gamma})^2$.
Following this rule, we find the following operator

\begin{eqnarray}
{\cal O}_{def}&=&
\Tr\left(\Phi^1 \Phi^1 \Phi^2 \Phi^2 \Phi^3 \Phi^3\right)+
\alpha\Tr\left(\Phi^1 \Phi^1 \Phi^2 \Phi^3 \Phi^2 \Phi^3\right)
\nonumber\\
&+&\alpha^2 \Tr\left(\Phi^1 \Phi^1 \Phi^2 \Phi^3 \Phi^3 \Phi^2\right)+
\alpha^2 \Tr\left(\Phi^1 \Phi^1 \Phi^3 \Phi^2 \Phi^2 \Phi^3\right)
\nonumber\\
&+&\alpha^3 \Tr\left(\Phi^1 \Phi^1 \Phi^3 \Phi^2 \Phi^3 \Phi^2\right)+
\alpha^4 \Tr\left(\Phi^1 \Phi^1 \Phi^3 \Phi^3 \Phi^2 \Phi^2\right)
\nonumber\\
&+&\alpha\Tr\left(\Phi^1 \Phi^2 \Phi^1 \Phi^2 \Phi^3 \Phi^3\right)+
\alpha^2 \Tr\left(\Phi^1 \Phi^2 \Phi^1 \Phi^3 \Phi^2 \Phi^3\right)
\nonumber\\
&+&\alpha^3\Tr\left(\Phi^1 \Phi^2 \Phi^1 \Phi^3 \Phi^3 \Phi^2\right)+
\alpha^2 \Tr\left(\Phi^1 \Phi^2 \Phi^2 \Phi^1 \Phi^3 \Phi^3\right)
\nonumber\\
&+&\alpha\Tr\left(\Phi^1 \Phi^2 \Phi^2 \Phi^3 \Phi^1 \Phi^3\right)+
\alpha\Tr\left(\Phi^1 \Phi^2 \Phi^3 \Phi^1 \Phi^2 \Phi^3\right)
\nonumber\\
&+&\alpha^2 \Tr\left(\Phi^1 \Phi^2 \Phi^3 \Phi^1 \Phi^3 \Phi^2\right)+
\alpha^2 \Tr\left(\Phi^1 \Phi^2 \Phi^3 \Phi^2 \Phi^1 \Phi^3\right)
\nonumber\\
&+&\alpha^3 \Tr\left(\Phi^1 \Phi^3 \Phi^1 \Phi^3 \Phi^2 \Phi^2\right)+
\alpha^3 \Tr\left(\Phi^1 \Phi^3 \Phi^2 \Phi^1 \Phi^3 \Phi^2\right).
\nonumber
\end{eqnarray}

\noindent
for $J=6$. Notice that when $\gamma=0$, this again reduces to a BPS operator
of the undeformed theory.
As a second example, here are the first few terms for the $J=9$
operator dual to the vacuum

\begin{eqnarray}
{\cal O}&=& \Tr \left( (\Phi^1)^3(\Phi^2)^3(\Phi^3)^3\right)+
\alpha\Tr \left((\Phi^1)^3(\Phi^2)^2\Phi^3\Phi^2(\Phi^3)^2\right)+
\alpha^2\Tr \left((\Phi^1)^3(\Phi^2)^2(\Phi^3)^2\Phi^2\Phi^3\right)
\nonumber\\
&+&\alpha^3\Tr \left((\Phi^1)^3(\Phi^2)^2(\Phi^3)^3\Phi^2\right)+...
\nonumber
\end{eqnarray}

\noindent
There are a total of 188 terms in the above sum.

One may worry that our prescription to obtain the operator dual to the vacuum is not
well defined. What is at stake here, is the fact that this 
prescription might be ambiguous. If there is more than one sequence of even and odd 
exchanges that will produce a particular word, we must check that each distinct
sequence of exchanges
assigns the same phase. In specific examples, we have checked that this is indeed 
the case.

\subsection{Operators Dual to Excited String Modes in the Undeformed Theory}

In this subsection, we set $\gamma=0$. Lets us consider the original pp-wave
limit of \cite{BMN}. Towards this end, imagine taking the pp-wave limit by boosting 
along the $\Phi^2$ direction (instead of along $\psi$). Define

\begin{equation}
\tilde {\cal O}_{(n)}=\Tr (\Phi^1(\Phi^2)^n\Phi^3 (\Phi^2)^{J-n}).
\label{loops}
\end{equation}

\noindent
which have two point functions

$$\langle \tilde{\cal O}_{(n)}(x_1)\bar{\tilde{\cal O}}_{(m)}(x_2)\rangle
=\delta_{mn}{N^{J+2} a^{J+2}\over |x_1-x_2|^{2J+4}}.$$

\noindent
To obtain the anomalous dimensions of the 
$\tilde {\cal O}_{(n)}$ we need to diagonalize $Q$ where

$$\langle\tilde {\cal O}_{(i)}(x_1)V_F(y)\bar{\tilde {\cal O}}_{(j)}(x_2)\rangle 
=Q_{ij}{N^{J+3}a^{J+4}\over |x_1-x_2|^{2J}|x_1-y|^4 |x_2-y|^4}.$$

\noindent
At leading order in $N$, we find

$$ Q=2\left[
\matrix{
3   &-2  &0   &0   &0   &\cdots &0   &0  &-1 \cr
-2  &4   &-2  &0   &0   &\cdots &0   &0  &0  \cr
0   &-2  &4   &-2  &0   &\cdots &0   &0  &0  \cr
0   &0   &-2  &4   &-2  &\cdots &0   &0  &0  \cr
:   &:   &:   &:   &:   &\cdots &:   &:  &:  \cr
0   &0   &0   &0   &0   &\cdots &-2  &4  &-2 \cr
-1  &0   &0   &0   &0   &\cdots  &0   &-2 &3}\right].$$

\noindent
The eigenvalues of $Q$ determine the anomalous dimensions of
$\tilde {\cal O}_{(n)}$. The eigenvectors of $Q$ determine the operators that
are dual to excited string modes.

Since the undeformed theory has an $SO(6)$ rotational invariance, the anomalous dimensions 
of the loops discussed above should agree with the anomalous dimensions of the loops obtained
in the pp-wave limit we are interested in. We will show that this is indeed the case.

If we look at the equation (\ref{loops}) one can think that the fields $\Phi^1$ define
a lattice, and that Yang-Mills interaction can be described in terms of the fields 
$\Phi^2$ and $\Phi^3$ ``hopping on this lattice". In our pp-wave limit, there isn't a
field which is singled out to play the r\^ole of a lattice. However, in analogy to
(\ref{loops}), define

$${\cal O}_{(n)}\equiv  C_{i_1 i_2 i_3 \cdots i_J}\Tr (
\Phi^1\Phi^{i_1}\cdots \Phi^{i_n}\Phi^3 \Phi^{i_{n+1}}\cdots \Phi^{i_J}).$$

The first thing we need to compute is the overlap
$\langle{\cal O}_{(l)}\bar{\cal O}_{(k)}\rangle .$
Introduce the notation (repeated indices summed as usual)

$$ C^2\equiv C_{i_1 i_2 \cdots i_J}C_{i_1 i_2 \cdots i_J}.$$

\noindent
The operator dual to the sigma model vacuum relevant for our pp-wave limit has an equal number
of $\Phi^1$s, $\Phi^2$s and $\Phi^3$s. This implies that (the exact location of the 
indices that are 1s or 2s or 3s is unimportant because $C$ is a symmetric tensor)

$$C_{1 i_2 \cdots i_J}C_{1 i_2 \cdots i_J}=
  C_{2 i_2 \cdots i_J}C_{2 i_2 \cdots i_J}=
  C_{3 i_2 \cdots i_J}C_{3 i_2 \cdots i_J}={1\over 3}C^2,$$
$$C_{1 2 \cdots i_J}C_{1 2 \cdots i_J}=
  C_{2 3 \cdots i_J}C_{2 3 \cdots i_J}=
  C_{3 1 \cdots i_J}C_{3 1 \cdots i_J}=AC^2,$$
$$C_{1 1 i_3 \cdots i_J}C_{1 1 i_3 \cdots i_J}=
   C_{2 2 i_3 \cdots i_J}C_{2 2 i_3 \cdots i_J}=
   C_{3 3 i_3 \cdots i_J}C_{3 3 i_3 \cdots i_J}=BC^2,$$

\noindent
where

$$ 6A+3B=1.$$

\noindent
It is now simple to argue that ($k\ne l$; in the second equation below,
$l$ is {\it not} summed)

$$\langle{\cal O}_{(l)}(x_1)\bar{\cal O}_{(k)}(x_2)\rangle = 
\left[(J-2)AC^2 + {2\over 3}C^2+\delta_{l,J-k}AC^2\right] N^{J+2}
{a^{J+2}\over |x_1-x_2|^{2J+4}} ,$$
$$\langle{\cal O}_{(l)}(x_1)\bar{\cal O}_{(l)}(x_2)\rangle = 
\left[ (J-1)AC^2 + C^2\right]N^{J+2}
{a^{J+2}\over |x_1-x_2|^{2J+4}}.$$

\noindent
At large $J$ we can write

$$\langle{\cal O}_{(l)}(x_1)\bar{\cal O}_{(k)}(x_2)\rangle = 
\left[(J-2)AC^2 + {2\over 3}C^2\right] N^{J+2}
{a^{J+2}\over |x_1-x_2|^{2J+4}} .$$

\noindent
Thus, ($k$ and $l$ unrestricted)

\begin{eqnarray}
\langle{\cal O}_{(l)}(x_1)\bar{\cal O}_{(k)}(x_2)\rangle 
&\equiv& M_{lk}{N^{J+2} a^{J+2}\over |x_1-x_2|^{2J+4}}
\nonumber\\
&=&\left[
\left((J-2)AC^2 +{2C^2\over 3}\right)L +(A+{1\over 3})C^2 I\right]_{lk} 
{N^{J+2} a^{J+2}\over |x_1-x_2|^{2J+4}},
\nonumber
\end{eqnarray}

\noindent
where $L$ is a matrix with a 1 in every single entry and $I$ is the identity matrix.
In what follows we will need the eigenvectors and eigenvalues of $M_{kl}$. There are $J$
eigenvectors that have the form (the first $n$ entries are 1s; $n=1,2,...,J$) 

$$|n\rangle ={1\over \sqrt{n^2+n}}
\left[\matrix{1\cr 1\cr :\cr 1\cr -n\cr 0\cr :\cr 0}\right].$$

\noindent
These have eigenvalue $(A+{1\over 3})C^2$. There is a single eigenvector of the form

$$ |J+1\rangle ={1\over\sqrt{J+1}}\left[\matrix{1\cr 1\cr :\cr :\cr 1}\right].$$

\noindent
This eigenvector has eigenvalue 
$(J+1)\left((J-2)AC^2 +{2C^2\over 3}\right) +(A+{1\over 3})C^2.$
These eigenvalues and eigenvectors can be used to define the new operators
${\cal K}_{(n)}$ that have a diagonal two point function. Explicitly we have

$${\cal K}_{(n)}={\langle n|_l {\cal O}_l\over\sqrt{\lambda_n}},$$

\noindent
and

$$\bar {\cal K}_{(n)}={\bar{\cal O}_l|n\rangle_l\over\sqrt{\lambda_n}}.$$

\noindent
These operators have two point function

$$\langle\bar {\cal K}_{(n)}(x_1){\cal K}_{(m)} (x_2)\rangle =
\delta_{mn}{N^{J+2} a^{J+2}\over |x_1-x_2|^{2J+4}}.$$

\noindent
To determine the anomalous dimensions for this set of operators at $O(g_{YM}^2)$ we compute

$$\langle{\cal O}_{(i)}(x_1)V_F(y)\bar{\cal O}_{(j)}(x_2)\rangle 
=H_{ik}M_{kj}{N^{J+3}a^{J+4}\over |x_1-x_2|^{2J}|x_1-y|^4 |x_2-y|^4}.$$

\noindent
At large $N$ we obtain

$$ H=2\left[\matrix{
3   &-2  &0   &0   &0   &\cdots &0   &0  &-1 \cr
-2  &4   &-2  &0   &0   &\cdots &0   &0  &0  \cr
0   &-2  &4   &-2  &0   &\cdots &0   &0  &0  \cr
0   &0   &-2  &4   &-2  &\cdots &0   &0  &0  \cr
:   &:   &:   &:   &:   &\cdots &:   &:  &:  \cr
0   &0   &0   &0   &0   &\cdots &-2  &4  &-2 \cr
-1  &0   &0   &0   &0   &\cdots  &0   &-2 &3}\right].$$

\noindent
Using this result, it is a simple matter to demonstrate

$$\langle{\cal K}_{(i)}(x_1)V_F(y)\bar{\cal K}_{(j)}(x_2)\rangle 
=K_{ij}{N^{J+3}a^{J+4}\over |x_1-x_2|^{2J}|x_1-y|^4 |x_2-y|^4},$$

\noindent
where

$$ K_{nm}=\sqrt{\lambda_m\over\lambda_n}\langle n|H|m\rangle .$$

\noindent
It is the eigenvalues of $K$ that determine the anomalous dimensions. The eigenvectors
of $K$ give the corresponding operators dual to excited string modes.
The prefactor $\sqrt{\lambda_m\over\lambda_n}$ differs from 1 only if $m=J+1$
or if $n=J+1$. As a consequence, noting that

$$\langle J+1|H=0=H|J+1\rangle,$$

\noindent
we see that we can write

$$ K_{nm}=\langle n|H|m\rangle .$$

\noindent
This implies that $K$ and $H$ are related by a unitary
transformation and hence we may as well solve the
eigenvalue problem for $H$. Since $Q$ and $H$ are identical matrices, this
demonstrates that the spectrum of our pp-wave limit agrees with the spectrum of the pp-wave
limit taken in \cite{BMN}, as expected from the rotational invariance of the background.
This agreement between the two computations gives us confidence
that we have indeed identified the operators dual to excited string states.

A few comments are in order. In our analysis, we have focused on $J+1$ operators. If
we write down the full set of operators with specific $U(1)_{\cal R}$ 
charge $J+2$ and $U(1)\times U(1)$ charge
equal to $(1,1)$ we find many more than just $J+1$ operators. Indeed, for
$J=6$ ($J=9$) we have kept only 7 (10) operators out of a possible 70 (1050) operators
with the correct quantum numbers. For $J=3$ and $J=6$ we have checked explicitly,
using the full set of loops,
that the $J+1$ BMN operators we have obtained by keeping only this subset of $J+1$ 
operators do indeed provide operators with a definite 
anomalous dimension at $O(g_{YM}^2)$. Further, we checked 
that the anomalous dimension we obtained 
agrees with the anomalous dimension obtained when the full class of operators is 
considered. 

This decoupling of a small set of nearly protected states has been used in 
both \cite{BMN} and \cite{Berenstein}. Understanding this decoupling directly
in the relevant quantum field theory is an important problem. The analysis of this 
section provides some insight into this decoupling in the field theory. The usual
argument\cite{BMN},\cite{Berenstein} involves taking a limit in which all states
that are not nearly protected have a very large energy and hence decouple. In this 
subsection we have seen that, at this order in perturbation theory, the potential 
coupling between the nearly protected states and other states vanishes.

\subsection{Operators Dual to Excited String Modes in the Deformed Theory}

In this section we will study operators dual to excited string modes for both
large $J$ and small $J$. This allows us to verify the $\gamma$ independence of
the large $J$ spectrum and further, that this is no longer the case at finite $J$.

Consider the large $J$ limit. First, we build the ``background" on which the 
impurities move. The background is built from an even number of $\Phi^1$, $\Phi^2$
and $\Phi^3$ fields. Start by selecting one of the Higgs fields from which the background 
is to be composed. Place a second background Higgs field to the left of this first one
and let it hop over the first, assigning phases for even and odd exchanges as in section
4.2. Place a third background Higgs field to the left of the two terms generated,
and let it hop all the way to the right, generating a total of 6 terms. Continue until
all background Higgs fields have been selected. As an example, if we wanted to build
the background out of one $\Phi^1$, one $\Phi^2$ and one $\Phi^3$, we would find go
through the following steps

\begin{eqnarray}
\nonumber
\Phi^1\to \Phi^2 \Phi^1 +e^{2\pi i\gamma}\Phi^1\Phi^2\to
\Phi^3\Phi^2 \Phi^1 &+&e^{2\pi i\gamma}\Phi^2\Phi^3 \Phi^1 +
\Phi^2\Phi^1 \Phi^3
+e^{2\pi i\gamma}\Phi^3\Phi^1\Phi^2\\
\nonumber
&+&\Phi^1\Phi^3\Phi^2
+e^{2\pi i\gamma}\Phi^1\Phi^2\Phi^3 .
\end{eqnarray}

\noindent
Selecting the background fields in a different order may change the
overall (and hence arbitrary) phase of the above operator. By building the operator
in this way, each exchange term we add by hand will be matched by an exchange
performed by the potential, with an {\it opposite sign} so that this indeed
builds a BPS state. This is not quite exact, because we did not consider the exchange
that will swap the last and first Higgs field. However, we expect that neglecting this 
exchange is justified in the leading order of a large $J$ expansion. Notice that if
the above operator is now traced, it will not in general reduce to the BPS 
state we identified in section 4.2. This can be traced back to our neglect of the
exchange of the first and last Higgs fields.

We will now describe how to build excited string states with two impurities. For the
impurities take $\Phi^1$ and $\Phi^3$. Let $\Phi^3$ hop
into the $n$th position using the same rules for hopping as above. The operator
obtained in this way is ${\cal O}^\gamma_{n}$. For each ${\cal O}_n^\gamma$
let $\Phi^1$ hop into the $m$th position. Call the resulting operator 
${\cal O}_{n,m}^\gamma$. Now define ($p=0,1,...,J$)

$${\cal O}^\gamma_{(p)}=\sum_{n,m}\Tr\Big({\cal O}_{n,m}^\gamma\Big)\delta_{m-n,p},$$

\noindent
where the delta function sets $m-n=p\, {\rm mod}\, J$.
It is now a simple task to show that

$$ \langle {\cal O}^\gamma_{(i)}(x_1)V_F(y)\bar{\cal O}^\gamma_{(j)}(x_2)\rangle
=H^\gamma_{ik}M^\gamma_{kj}{N^{J+3}a^{J+4}\over |x_1-x_2|^{2J}|x_1-y|^4|x_2-y|^4},$$

\noindent
where

$$ \langle {\cal O}^\gamma_{(i)}(x_1)\bar{\cal O}^\gamma_{(j)}(x_2)\rangle
=M^\gamma_{ij}{N^{J+2}a^{J+2}\over |x_1-x_2|^{2J+4}}.$$

\noindent
When computing these correlators, we sum over all contractions except the contractions
involving the fields that were at the endpoints of ${\cal O}_{n,m}^\gamma$; this
should give the correct answer in the large $J$ limit.
In the above, we have

$$ H^\gamma_{ik}=8\delta_{ik}-4\delta_{i+1 k}-4\delta_{i k+1},$$

\noindent
This looks the same as $H$ of section 4.2 except that we don't have the
-1 elements in $H_{0,J}$ and $H_{J,0}$. In the large $J$ limit, we expect (and have verified
numerically) that the precise 
details of these terms are unimportant, so that $H^\gamma$
has the same spectrum as $H$ in section 4.3. Our proposal for the BMN operators is then
to build them using the eigenvectors of $H^\gamma$. Thus, we see that the spectrum
of anomalous dimensions coincides with the spectrum of anomalous dimensions of 
the undeformed ($\gamma=0$) theory, in perfect agreement with the string theory 
prediction.

This conclusion assumes that the eigenvalues of $H^\gamma$ determine the anomalous
dimensions of the operators we consider. In the undeformed case we were able to argue
that this is indeed the case by studying the eigenvalues and eigenvectors of $M_{ij}$.
To prove this assumption in the deformed case, we would need to provide the
corresponding study for $M_{ij}^\gamma$. Although our assumption seems reasonable, we
have not proved that it is indeed correct.

We now consider the small $J$ limit. For small values of $J$, 
we can work with the full set of loops 
${\cal O}_i$ that have $U(1)\times U(1)$ charge (1,1) and $U(1)_{\cal R}$ charge 
$J+2$. The ${\cal O}_i$ are chosen to have two point function

$$ \langle {\cal O}_i\bar{\cal O}_j\rangle\propto\delta_{ij}$$

\noindent
at large $N$ and $O(g_{YM}^0)$. Computing correlators of the form

$$\langle{\cal O}_{i}(x_1)V_F(y)\bar{\cal O}_{j}(x_2)\rangle 
=T_{ij}{N^{J+3}a^{J+4}\over |x_1-x_2|^{2J}|x_1-y|^4 |x_2-y|^4},$$

\noindent
the matrix $T$ determines the operators with a definite anomalous dimension and the
anomalous dimension itself, to $O(g_{YM}^2)$. We find that for $\gamma=0.1$ and $J=3$
the smallest eigenvalue of $T$ is 0.07843... and for $J=6$, the smallest eigenvalue is
0.04124... The string theory prediction of section 3, which corresponds to infinite $J$,
is that this smallest eigenvalue should be zero. The fact that the smallest eigenvalue
is non-zero is a clear indication that we can't compare our finite $J$ field theory 
results with the string theory results. We have also developed an expansion for $T$
in terms of $\gamma$. It is then possible (using the results of the appendix) to
develop a perturbative expansion (treating $\gamma$ as a small number) for the anomalous
dimension. We find that the $O(\gamma )$ term vanishes. 
One could in principle develop this perturbation to even 
higher orders. Reproducing this perturbation series directly in the string theory would 
be an interesting exercise.

\section{Summary}
We have taken a pp-wave limit of the Lunin-Maldacena background. The resulting
geometry is that of a homogeneous plane wave. The spectrum of the string
is independent of the deformation parameter $\gamma$. In the 
dual gauge theory, we have argued that for the class of operators we consider, at
$O(g_{YM}^2)$ and at leading order in $N$, all contributions to the anomalous dimension
come from $F$-terms. We have identified the operator in the deformed super Yang-Mills 
which is dual to the sigma model vacuum state.  For the undeformed theory, we have
been able to identify a set of operators dual to excited string modes. Further, these
operators are a small fraction of the total number of operators with the correct
quantum numbers to participate. This sheds some light on the important issue of
decoupling a small set of nearly protected states\cite{BMN},\cite{Berenstein}.
For the deformed theory, we proposed a set of operators  dual to excited string modes, 
for large $J$. The anomalous dimensions of these operators are independent of $\gamma$
in perfect agreement with the string theory spectrum. 
For finite $J$, at order $g_{YM}^2$, the anomalous dimensions we have computed do
depend on $\gamma$. It would be interesting to reproduce this dependence in the
string theory, presumably by adding ${1\over J}$ corrections.

$$ $$

\noindent
{\it Acknowledgements:} 
We would like to thank T. Mateos for pointing out an error in our string
spectrum which appeared in a previous draft.
The work of RdMK, JS and MS is supported by NRF grant number Gun 2047219.
JM is supported by an overseas postdoctoral fellowship of the NRF (South Africa).

$$ $$

\appendix{\bf Appendix: Eigenvalue Problem}

In this appendix we solve the eigenvalue problem of the operator $H$ introduced in
section 4.3. Denoting the components of the eigenvectors

$$ H|i\rangle=\lambda_i|i\rangle ,$$

\noindent
by

$$ |i\rangle =\left[\matrix{v_0\cr v_1\cr :\cr v_{J-1}\cr v_J}\right],$$

\noindent
we have

\begin{equation}
-4v_{n-1}+8v_n -4v_{n+1}=\lambda v_n
\label{evone}
\end{equation}

\noindent
for $1\le n\le J-1$ and

\begin{equation}
3v_0-2v_1 -v_J=\lambda v_0\qquad 3v_J-2v_{J-1}-v_0=\lambda v_J.
\label{evtwo}
\end{equation}

\noindent
Make the ansatz

$$ v_n=Ae^{ikn}+Be^{-ikn}.$$

\noindent
Then, (\ref{evone}) implies $\lambda =8-8\cos (k)$, and (\ref{evtwo}) implies that the
allowed values of $k$ solve

$$ {\rm Imag}\left[(\lambda-3+2e^{-ik}+e^{-iJk})
(3e^{ikJ}-2e^{i(J-1)k}-1-\lambda e^{iJk})\right]=0,$$

\noindent
where ${\rm Imag}$ stands for the imaginary part. It is now a simple exercise to determine
$A$ in terms of $B$ using (\ref{evtwo}). $B$ is determined by 
the normalization of the eigenvector.


\begin{thebibliography}{30}
\parskip-2pt

\bibitem{AdSCFT}
J. Maldacena, ``The large N limit of superconformal field theories and supergravity,"
Adv. Theor. Math. Phys. {\bf 2} 231 (1998), {\tt hep-th/9711200};\\
S. Gubser, I.R. Klebanov and A.M. Polyakov, ``Gauge Theory Correlators from Non-critical
String Theory," Phys. Lett.{\bf  B428} (1998) 105, {\tt hep-th/9802109};\\
E. Witten, ``Anti-de Sitter Space and Holography,"
Adv. Theor. Math. Phys. {\bf 2} (1998) 253, {\tt hep-th/9802150};\\
Ofer Aharony, Steven S. Gubser, Juan M. Maldacena, Hirosi Ooguri and Yaron Oz,
``Large N Field Theories, String Theory and Gravity,"
Phys. Rept. {\bf 323} (2000) 183, {\tt hep-th/9905111}. 

\bibitem{BMN}
D. Berenstein, J.M. Maldacena and H. Nastase, 
``Strings in Flat Space and pp-waves from N=4 super Yang-Mills,"
JHEP {\bf 0204}, 013 (2002), {\tt hep-th/0202021}.

\bibitem{LM}
O. Lunin and J.M. Maldacena, ``Deforming Field Theories with $U(1)\times U(1)$ global 
symmetry and their gravity duals," {\tt hep-th/0502086}.

\bibitem{GKP}
S.S. Gubser, I.R. Klebanov and A.M. Polyakov, ``A Semi-classical Limit of the Gauge/String
Correspondence," Nucl. Phys. {\bf B636}, 99 (2002), {\tt hep-th/0204051};\\
S. Frolov and A.A. Tseytlin, ``Multi-spin string solutions in AdS$_5\times$S$^5$ and beyond,"
Nucl. Phys. {\bf B668}, 77 (2003), {\tt hep-th/0304255};\\
A.A. Tseytlin, ``Semiclassical Quantization of Superstrings: AdS$_5\times$S$^5$ and beyond,"
Int. J. Mod. Phys. {\bf A18}, 981 (2003), {\tt hep-th/0209116};\\
A.A. Tseytlin, ``Spinning Strings and AdS/CFT Duality," {\tt hep-th/0311139}.

\bibitem{FRT}
S.A. Frolov, R. Roiban and A.A. Tseytlin, ``Gauge-string duality for superconformal
deformations of $N=4$ Super Yang-Mills Theory," {\tt hep-th/0503192}.

\bibitem{F}
S.A. Frolov, ``Lax Pair for Strings in Lunin-Maldacena Background," {\tt hep-th/0503201}.

\bibitem{Beisert}
N. Beisert and R. Roiban, ``Beauty and the Twist: The Bethe Ansatz for Twisted
${\cal N}=4$ SYM," {\tt hep-th/0505187}.

\bibitem{BK}
J. Gomis and H. Ooguri, ``Penrose Limits of N=1 Gauge Theories," Nucl. Phys.
{\bf B635} (2002) 106, {\tt hep-th/0202157};\\
Dominic Brecher, Clifford V. Johnson, Kenneth J. Lovis and Robert C. Myers,
``Penrose Limits, Deformed PP-waves and the String Duals of N=1 Large N Gauge Theory," 
{\bf JHEP} 0210:008, 2002 {\tt hep-th/0206045};\\ 
H. Dimov, V. Filev, R.C. Rashkov and K.S. Viswanathan, ``Semiclassical Quantization of
Rotating Strings in Pilch-Warner Geometry," Phys. Rev. {\bf D68} 066010, 2003, 
{\tt hep-th/0304035};
S. Benvenuti and M. Kruczenski, ``Semiclassical Strings in Sasaki-Einstein Manifolds
and Long Operators in N=1 Gauge Theories," {\tt hep-th/0505046};\\
S. Benvenuti and M. Kruczenski, ``From Sasaki-Einstein Spaces to quives via BPS
geodesics: Lpqr," {\tt hep-th/0505206}.

\bibitem{GN}
U. Gursoy and C. Nunez, ``Dipole Deformations of N=1 SYM and Supergravity Backgrounds
with $U(1)\times U(1)$ Global Symmetry," {\tt hep-th/0505100}.

\bibitem{CA}
C. Ahn and J.F. Vazquez-Portiz, ``Marginal Deformations with $U(1)^3$ Global Symmetry,"
{\tt hep-th/0505168}.

\bibitem{Lee}
J.P. Gauntlett, S. Lee, T Mateos and D. Waldram, ``Marginal Deformations of Field Theories
with AdS$_4$ Duals," {\tt hep-th/0505207}.

\bibitem{NP}
V. Niarchos and N. Prezas, ``BMN Operators for ${\cal N}=1$ superconformal Yang-Mills
theories and associated string backgrounds," JHEP {\bf 0306}, 015 (2003),
{\tt hep-th 0212111}.

\bibitem{HPP}
G. Papadopoulos, J.G. Russo and A.A. Tseytlin, ``Solvable Model of Strings in a time
dependent plane-wave background," Class. Quant. Grav. {\bf 20}, 969 (2003), 
{\tt hep-th/0211289};\\
M. Blau and M. O'Loughlin, ``Homogeneous Plane Waves," Nucl. Phys. {\bf B654}, 135 (2003), 
{\tt hep-th/0212135}.

\bibitem{SHPP}
R.R. Metsaev, ``Type IIB Green-Schwarz Superstring in Plane Wave Ramond-Ramond Background,"
Nucl. Phys. {\bf B625} 70 (2002), {\tt hep-th/0112044};\\
R.R. Metsaev and A.A. Tseytlin, ``Exactly Solvable Model of Superstring in Plane Wave 
Ramond-Ramond Background," Phys. Rev. {\bf D65} 126004 (2002), {\tt hep-th/0202109};\\
J.G. Russo and A.A. Tseytlin, ``On Solvable Models of Type IIB Superstring in 
NS-NS and R-R Plane Wave Backgrounds," JHEP {\bf 0204} 021 (2002), {\tt hep-th/0202179};\\
M. Blau, M. O'Loughlin, G. Papadopoulos and A.A. Tseytlin, ``Solvable Models of Strings in
Homogeneous Plane Wave Backgrounds," Nucl. Phys. {\bf B673}, 57 (2003), {\tt hep-th/0304198.}

\bibitem{Toni}
T. Mateos, ``Marginal deformations of N=4 SYM and Penrose limits with continuum spectrum,"
{\tt hep-th/0505243}.

\bibitem{Motl}
N.R. Constable, D.Z. Freedman, M. Headrick, S. Minwalla, L. Motl, A. Postnikov and W. Skiba,
``PP-wave String Interactions from perturbative Yang-Mills Theory,"
{\tt hep-th/0205089}.

\bibitem{Penati}
S. Penati, A. Santambrogio and Daniela Zanon, ``Two point functions
of Chiral Operators in N=4 SYM at order $g^4$," JHEP {\bf 9912}
006,(1999), {\tt hep-th/9910197};\\
Silvia Penati, Alberto Santambrogio and Daniela Zanon, ``More on
Correlators and Contact Terms in N=4 SYM at order $g^4$," 
Nucl. Phys. {\bf B593} 651-670 (2001), {\tt hep-th/0005223};\\
Silvia Penati and Alberto Santambrogio, ``Superspace Approach to
Anomalous Dimensions in N=4 SYM," Nucl. Phys. {\bf B614} 367-387
(2001) {\tt hep-th/0107071};\\
Alberto Santambrogio and Daniela Zanon, ``Exact Anomalous Dimensions
of N=4 Yang-Mills Operators with Large R Charge," Phys. Lett. {\bf B545} 425-429 (2002) {\tt hep-th/0206079}.

\bibitem{HFS}
E. D'Hoker, D.Z. Freedman and W.Skiba, ``Field Theory tests for correlators in the 
AdS/CFT Correspondence," Phys. Rev. {\bf D59} 045008 (1999), {\tt hep-th/9807098}.

\bibitem{Berenstein}
D. Berenstein, {\it A Toy Model for the AdS/CFT Correspondence,} {\tt hep-th/0403110}

\bibitem{DHK}
N. Dorey, T.J. Hollowood and S.P. Kumar, ``$S$ Duality of the Leigh-Strassler Deformation
via Matrix Models," 
JHEP {\bf 0212} 003 (2002), {\tt hep-th/0210239}.


\end{thebibliography}
\end{document}